# Topological Insulator-Based van der Waals Heterostructures for Effective Control of Massless and Massive Dirac Fermions

Su Kong Chong[1], Kyu Bum Han[2], Akira Nagaoka[2,3], Ryuichi Tsuchikawa[1], Renlong Liu[4,5], Haoliang Liu[1], Z. Valy Vardeny[1], Dmytro A. Pesin[1], Changgu Lee[4,5], Taylor D. Sparks[2] and Vikram V. Deshpande[1]*

[1]Department of Physics and Astronomy, University of Utah, Salt Lake City, Utah 84112 USA

[2]Department of Materials Science and Engineering, University of Utah, Salt Lake City, Utah 84112 USA

[3]Department of Materials Science and Engineering, Kyoto University, Kyoto 606-8501, Japan

[4]Department of Mechanical Engineering, Sungkyunkwan University, 2066, Seobu-ro, Jangan-gu, Suwon, Gyeonggi, 16419, Republic of Korea

[5]SKKU Advanced Institute of Nanotechnology, Sungkyunkwan University, 2066, Seobu-ro, Jangan-gu, Suwon, Gyeonggi, 16419, Republic of Korea

*Corresponding author: vdesh@physics.utah.edu

**Three dimensional (3D) topological insulators (TIs) are an important class of materials with applications in electronics, spintronics and quantum computing. With the recent development of truly bulk insulating 3D TIs, it has become possible to realize surface dominated phenomena in electrical transport measurements e.g. the quantum Hall (QH) effect of massless Dirac fermions in topological surface states (TSS). However, to realize more advanced devices and phenomena, there is a need for a platform to tune the TSS or modify them e.g. gap them by proximity with magnetic insulators, in a clean manner. Here we introduce van der Waals (vdW) heterostructures in the form of topological**

**insulator/insulator/graphite to effectively control chemical potential of the TSS. Two types of gate dielectrics, normal insulator hexagonal boron nitride (hBN) and ferromagnetic insulator $Cr_2Ge_2Te_6$ (CGT) are utilized to tune charge density of TSS in the quaternary TI $BiSbTeSe_2$. hBN/graphite gating in the QH regime shows improved quantization of TSS by suppression of magnetoconductivity of massless Dirac fermions. CGT/graphite gating of massive Dirac fermions in the QH regime yields half-quantized Hall conductance steps and a measure of the Dirac gap. Our work shows the promise of the vdW platform in creating advanced high-quality TI-based devices.**

The discovery of the three-dimensional topological insulator (3D TI) as a quantum state of matter has expanded the family of Dirac materials to higher dimensions[1,2]. Quaternary Bi-based TI compounds of the form $Bi_{2-x}Sb_xTe_{3-y}Se_y$ exhibit a large bulk band gap and more intrinsic bulk conduction which allow probing of the surface states[3]. Integer quantum Hall effect (QHE) in magnetotransport of 3D TIs provides a strong signature of the topological surface states (TSS)[4,5,6,7,8,9] and has been observed at a strong external magnetic field in $BiSbTeSe_2$ (BSTS) crystals[5,6,8,9]. The quantization of Hall conductivity in units of $(n+½)$ $e^2/h$, where $n$ is the (integer) Landau level index of the TSS, is due to two unique characteristics of TSS, namely spin non-degeneracy and $\pi$ Berry's phase[5,6,10]. Yet in experiment[5,6,7,8,9], two surfaces combine to give overall integer quantization and half-quantized conductivity values have not been realized controllably.

Realization of exotic effects in TSS requires fine control of the chemical potential near the Dirac point[11,12]. This can be achieved effectively by performing electrostatic gating using the highly configurable vdW platform[13]. The typical gate dielectric, $SiO_2$, suffers from electron scattering due to charged surface impurities[14]. The two-dimensional (2D) insulator hexagonal boron nitride (hBN) serves as a better gate dielectric owing to its relatively inert, surface-charge-free and atomically-flat surface[15]. The substrate induced disorder due to hBN has been measured in graphene to be an order magnitude smaller than from $SiO_2$[16]. Another interesting set of candidates are the ferromagnetic insulators (FMI) as they can possibly break time-reversal symmetry (TRS) on the TSS by exchange coupling proximity effect[17]. The layered chalcogenide $Cr_2Ge_2Te_6$ (CGT) can couple well with Bi-based TIs as it exhibits similar crystal structure and small lattice mismatch from the Bi-based TI[18]. Moreover, the out-of-plane magnetization in CGT can provide a stronger ferromagnetic exchange coupling to the TI surface than most oxide and suphide-based in-plane magnetization FMIs, such as YIG and EuS. More recently, graphite (Gr) gates have been used to reduce charge inhomogeneity in case of graphene devices[19]. Despite these advantages, 2D materials have not been widely explored in conjunction with TIs. In this work, we present BSTS-based vdW heterostructures in TI/insulator/graphite configuration, using hBN and FMIs as gate dielectrics and graphite as the gate, to achieve and control high-quality TI surfaces.

Device fabrication process by mechanical transfer technique is illustrated in Fig. 1(a). Fig. 1(b) shows the optical image of a typical dry transferred BSTS/hBN/Gr device (the transfer process of each layer is shown in Supplementary Fig. S1). To effectively induce charges on the surface of the BSTS, our devices are typically constructed from ≤20 nm hBN as dielectric and Gr layer as gate electrode. The transfer is carried out carefully at a temperature close to the vaporization temperature of water (~95°C) to avoid the water vapors trapping between hBN and Gr layers, which can adversely affect electrical transport in BSTS (Supplementary Fig. S2). Atomic force microscopy image as shown in Fig. 1(c) reveals the relatively flat and clean vdW assembly of the BSTS/hBN/Gr device. A step height trace over the device shows the thicknesses of each layer.

The longitudinal resistance ($R_{xx}$) of the BSTS as a function of topgate voltage applied through the Gr electrode is shown in Fig. 1(d). The BSTS device exhibits ambipolar transport with resistance maximum of ~9.3 kΩ located at charge neutrality point at $V_g$= -3.3V, indicating the top surface is electron-doped. The ambipolar resistance is a signature of surface transport in a TI. The induced charge density on the top BSTS surface is calculated as n= $C_g|V_g–V_D|$, where $C_g$ is the gate capacitance (≈ 150 nF/cm$^2$ for 18 nm hBN). This configuration allows us to induce surface charge density >5×10$^{12}$ /cm$^2$ in electron conduction. In the high density regime, the $R_{xx}$ falls sharply, resulting in a change of ~7 and ~4 kΩ in electron and hole conduction regions, respectively. The significant resistance change of the BSTS indicates an effective field-effect gate tuning due to good coupling of the 2D materials.

Transmission electron microscopy (TEM) was utilized to investigate the interfaces of the heterostructures. Cross-sectional TEM micrograph of a representative BSTS/hBN/Gr stack is shown in Fig. 2(a), where the layers are clearly revealed by contrast. Atomically sharp interfaces are formed at the boundaries, signifying good interface quality between the layers. This confirms the formation of highly ordered heterostructures using our dry transfer method. The quintuple layer structure of the BSTS is visibly resolved in the high resolution TEM micrograph recorded at the interface of BSTS and hBN (Fig. 2(b)). The lattice spacings of hBN and BSTS quintuple layers are determined from the contrast line profile taken across the adjacent lattice planes (Fig. 2(c)), which match well with their c-axis plane spacing. EDX elemental mapping analyses were performed (Fig. 2(d)) to confirm the cleanliness of each layer.

The high quality of our BSTS/hBN/Gr devices is confirmed by magnetotransport measurements. Our device configuration allows us to apply a topgate voltage ($V_{tg}$) and backgate voltage ($V_{bg}$) through the hBN and SiO$_2$ dielectrics, respectively, to individually modulate the charge densities on top and bottom surfaces. Fig. 3(a) shows the 2D color map of the longitudinal conductivity ($\sigma_{xx}$) of a BSTS dual-gated device as a function of dual-gate voltages measured at zero magnetic field. The Dirac fermions in top and bottom surface states are tuned independently to the chemical potential near the (top and bottom) Dirac cones. The conductivity minima of the top and bottom surfaces intersect at $V_{tg}$= -3.3V and $V_{bg}$= +3V, which corresponds to the overall charge neutrality point resulting from the top and bottom TSS combined. The color map can be divided into four quadrants with hole-hole (h-h), electron-electron (e-e), hole-electron (h-e) and electron-hole (e-h) conduction regions for charge carriers in top-bottom TI surfaces.

With an external magnetic field of 9 T applied perpendicular to the device, the $\sigma_{xx}$ (Fig. 3(b)) develops into four well-resolved quadrants (framed by dashed white line) with uniform rectangular areas forming near the Dirac points of the top and bottom surfaces. In both h-h and e-e conduction regions, the $\sigma_{xx}$ reaches a minimum value < 0.01 $e^2/h$. Vanishing of $\sigma_{xx}$ is an indication of the Landau level (LL) formation in the TSS. This is confirmed by the corresponding quantization in Hall resistance ($R_{xy}$) in the same parameter spaces as shown in the color map in Fig. 3(c). The $R_{xy}$ forms perfectly symmetric plateaus saturating ~-25 k$\Omega$ (~+25 k$\Omega$) in the h-h (e-e) conduction region. The $R_{xy}$ plot along the dashed red line (*i*) in Fig. 3(c) is extracted by plotting it as a function of sum of the top and bottom surface charge densities (red curve in Fig. 3(e)), where the quantized $R_{xy}$= ±h/$ve^2$ are labelled. Conversely, the $\sigma_{xx}$ develops oppositely in the h-e and e-h conduction regions. Instead of vanishing, the $R_{xx}$ enhances toward maximization (refer to Fig. 4(a) for the $R_{xx}$ dual-gates map) in conjunction with suppression in $R_{xy}$ in the region. Line profile of $R_{xy}$ across the e-h and h-e regions (dashed black line (*ii*) in Fig. 3(c)) shows two average-to-zero (-0 and +0) plateaus in the low-density regime, as plotted in Fig. 3(e). These dissipationless and dissipative QH states have been observed in the literature, albeit at higher magnetic fields[5,6,9].

To further analyze the QH states, we plot the Hall conductivity ($\sigma_{xy}$) as a function of charge densities induced in top and bottom surfaces in color scale as depicted in Fig. 3(d). The rectangular $\sigma_{xy}$ plateaus correspond to the addition of LL filling factors from the top and bottom surfaces with $v_t = n_t + ½$ and $v_b = n_b + ½$,[5,6,7,9] where $n_t$ and $n_b$ are the top/bottom surface LL indices and the ½ is a consequence of the π Berry's phase. The corresponding ($v_t$, $v_b$) are indexed in the color plot. The integer QHE with LL filling factors ($v = v_t + v_b$) of $v = 0, ±1$ and $±2$ is resolved in the figure. In addition, we compare the quantization on the top and bottom surfaces by extracting the $\sigma_{xx}$ and $\sigma_{xy}$ with $V_{tg}$ ($V_{bg}$) swept at $V_{bg}$ ($V_{tg}$) fixed to the bottom (top) Dirac point. Fig. 3(e) illustrates the renormalization group (RG) flow[20] ($\sigma_{xy}$ versus $\sigma_{xx}$) plots of the top and bottom quantized TSS. As the top and bottom surfaces are tuned separately by hBN and SiO$_2$ dielectric, the quantized states are determined solely by the quality of the dielectric materials. The $\sigma_{xx}$ of the QH states for the surface with hBN dielectric is suppressed to a greater extent, and the $\sigma_{xy}$ is closer to the quantized value, as compared to the surface with SiO$_2$ dielectric. This is indicative of better development of the QH state using hBN in comparison with the higher disorder SiO$_2$ dielectric. We have observed qualitatively similar results in six heterostructure devices.

The realization of the QHE in BSTS/hBN/Gr vdW heterostructures at lower magnetic fields than previous reported works demonstrates the high quality of our devices and effectiveness of the vdW platform for TIs. We further study electrical transport in BSTS interfaced with a ferromagnet by replacing the normal insulator hBN with a 2D FMI. CGT appears to be an appropriate FMI candidate as it can be mechanically exfoliated and integrated with BSTS devices. Kerr rotation of circularly polarized light from the CGT flakes measured using Sagnac interferometry confirms the ferromagnetic phase at temperatures below 60 K (Fig. S4 and S5). Temperature dependent resistance plot (Fig. S4(a)) shows the insulating behavior of CGT with resistance exceed GΩ at cryogenic temperatures.

In theory, the TSS in TIs can be gapped by proximity coupling to the FMI. Thus, electrostatic gating using CGT as dielectric to tune the chemical potential of the gapped TSS is highly desirable. We have created dual-gated BSTS/CGT/Gr devices and acheived ambipolar transport using CGT dielectric and Gr gate (Fig. S6(a)). Similar to BSTS/hBN/Gr devices, the $R_{xx}$ can be tuned into four quadrants (Fig. S6(b)) and magnetotransport develops into well-formed QH states in an external magnetic field of 9 T. However, in contrast to typical BSTS/hBN/Gr devices (Fig 4(a)), the four-quadrant QH plateaus in BSTS/CGT/Gr devices at the global charge neutrality point (Fig. 4(b)) is highly asymmetric, with the quantization of the top surface state between $n_t$=-1 and $n_t$=0 LLs extended over a longer range in $V_{tg}$ compared to the plateau between $n_t$=0 and $n_t$=1. We understand this asymmetry as a gap opening due to the TRS-broken surface state by the magnetic CGT. The corresponding energy band diagrams with the LLs formed in the linear and gapped Dirac dispersions are illustrated. An equal LL energy spacing between $n_t$= -1, 0 and +1 is observed when the chemical potential is tuned across the Dirac point by hBN/Gr gating. In contrast, for BSTS/CGT, the magnetic gap opened at the Dirac point causes an upshift[21,22] in $n_t$= 0 LL which resides at the bottom of the conduction band[8,23]. As a result, the band separation between $n_t$= -1 and 0 LLs is larger than the $n_t$= 0 and +1 LLs due to the presence of magnetic ordering at the BSTS/CGT interface. This observation is consistent with theoretical calculations and experimental reports for magnetic-doped TI[21,22], magnetic-doped/undoped TI bilayer[23] and Co-decorated/BSTS[8] systems. For clean BSTS/CGT interfaces (no impurities or trap states), the magnetic-gap size ($\Delta$) can be semi-quantitatively estimated from the relation, $\Delta = E_{0,-1} - E_{1,0}$, where $E_{0,-1}$ ($E_{1,0}$) is the LL spacing of $E_0$ & $E_{-1}$ ($E_1$ & $E_0$)[22]. $\Delta$ is evaluated from the asymmetry in LL spacing to be ~26 meV at 9T, much higher than Zeeman spin-splitting ~ 1-3 meV[24], indicating the magnetic mass origin of this gap.

The consequences of the Dirac gap are even more pronounced in the value of the quantization. Fig. 4(c) (inset) displays a map of the $\sigma_{xy}$ plateaus developed near the global charge neutrality point of BSTS/CGT/Gr. Fig. 4(c) shows the line profiles of $\sigma_{xx}$ and $\sigma_{xy}$ of the top surface as a function of $V_{tg}$ (with $V_{bg}$ tuned to hole conduction in the bottom surface state). The corresponding plots for the bottom surface are also included for comparison. Surprisingly, $\sigma_{xy}$ of the magnetized top surface is quantized in steps of $e^2/2h$. The corresponding RG flow plot in Fig. 4(d) clearly illustrates half-quantization steps with total $\nu= 0$, $-\frac{1}{2}$, and $-1$. In comparison, fixing the top surface state in the hole conduction region, the bottom Dirac surface state (gated using the $SiO_2$/Si backgate) shows the normal QHE with integer $e^2/h$ steps, similar to the observation in Fig. 3(f). Figs. S7-10 show data for both the Dirac gap and half-quantization in additional devices. A single half-quantized conductance value was observed recently in cobalt-decorated BSTS system[8]. However their TI surface could not be gated, preventing tuning of this state.

It is well-known that the existence of a true Hall plateau with half-quantized Hall conductivity is forbidden in a non-interacting system, as it violates gauge invariance. It is unlikely that electronic correlations play any significant role in our system, hence we are not dealing with a true half-quantized Hall plateau. That this is so is signaled by the fact that the feature in $\sigma_{xy}$ is accompanied by a substantial value of $\sigma_{xx}$ (a fraction of $e^2/h$), much larger than that for the true Hall plateau at $\sigma_{xy}= -e^2/h$ (see Fig. 4(c)). The quantitative description of the transport properties of our sample necessarily takes into account both 2D bulk transport on topological surfaces, and edge conduction on sample's sides. The edge transport in the quantum anomalous Hall effect on topological surfaces is complicated by the existence of chiral and non-chiral edge states[25]. In our system, further complexity is added by the proximity to a ferromagnet (CGT), which strongly modifies the band structure of surface and 3D bulk states, and the importance of the disorder, as signaled by the fact that only a few lowest Hall plateaus corresponding to the largest cyclotron gaps are developed. However, at the qualitative level, it appears clear that the quasi-plateau at $\sigma_{xy} \approx -e^2/2h$ develops close to the $\sigma_{xy}^t = -e^2/2h \rightarrow +e^2/2h$ transition at the top surface, which corresponds to zeroth Landau level of the top surface crossing the chemical potential of the system, while the bottom surface is kept at $\sigma_{xy}^b = -e^2/2h$. The fairly constant value of $\sigma_{xx}$ in the quasi-plateau region points either to the importance of edge transport in that regime, or to the fact that (localized) 3D bulk states participate in accommodation of electric charge induced by the gate electrodes, which would further promote zeroth Landau level pinning to the chemical potential, and stabilization of the quasi-plateau.

In conclusion, we studied the low temperature magnetoelectric transport of BSTS devices constructed using vdW heterostructures BSTS/hBN/Gr. By electrostatic gating, we achieved fully developed LL filling factor $\nu= 0, \pm 1$, and developing $\nu= \pm 2$ and $\pm 3$ QH states at magnetic fields below 9 T. This was attributed to the effective gate tuning using the vdW platform and our clean device fabrication process. The vdW epitaxial hBN/Gr gate led to better development of the QHE than $SiO_2$/Si gate, as indicated by a larger suppression of $\sigma_{xx}$ and better quantization in $\sigma_{xy}$. We also showed the ability to use the vdW platform for fabricating heterostructures with CGT/Gr gate, which allowed access and tuning of gapped TSS. The observation of half-quantized Hall conductance steps for the first time in a system where electron-electron correlations do not play a significant role provides a clear manifestations of massive Dirac fermion physics on the magnetized TI surface. The effective control of the TI surface states using 2D layered materials pave the way for epitaxial stacking, promising improved quality and tunability of TI devices.

## Methods

**Single crystal growth.** BSTS single crystals were grown using a vertical Bridgman method[5,26]. Bi, Sb, Te and Se metal pieces (Sigma-Aldrich, purity 99.999%) with molar ratio of 1:1:1:2 were sealed in a 12 mm quartz tube in a vacuum environment. The metal pieces were stirred at 850°C in a muffle furnace to form a uniform mixture polycrystalline BSTS. Three zones vertical Bridgman furnace was used for the growth of BSTS single crystals. The temperatures of zone A, B and C were set to 550°C, 800°C and 620°C, respectively, where the zone A is located at the top. The growth was carried out by placing the ampoule vertically 10 cm above the center of zone A, and passed down the three hot-zones at an extremely slow rate (6 mm/day). More details on the setup and synthesis methods can be found in the references[26]. CGT single crystals were synthesized by the flux method with the starting materials of high purity elemental Cr (99.5%), Ge (>99.99%), and Te (>99.8%) powders from Sigma-Aldrich[27]. A mixture of these elements with a molar ratio of Cr: Ge: Te equal to 10: 10: 76.5, was vacuum-sealed in a quartz ample. Then the ampule was vertically placed into a muffle furnace, and kept at 1000°C for 48h, followed by very slow cooling down to 400°C. The resultant single crystal platelets were taken out from the ampule and annealed at 450°C for 2 days to sublimate the residual Tellurium.

**Characterization techniques.** The morphology of the exfoliated BSTS crystal flakes were characterized by a Leica optical microscopy. The thickness of the samples was measured using a Bruker Dimension Icon atomic force microscopy. The scanning was performed in ScanAsyst mode using a silicon tip on nitride cantilever (spring constant= 0.4 N/m). Microstructures and elemental analyses were done using a JEOL JEM-2800 scanning transmission electron microscopy (STEM). TEM cross-section specimens were prepared using a FEI Helios NanoLab™ 650 focused ion beam dual beam microscope. The three layers samples were coated with about 100 nm of amorphous carbon, followed by a thick Pt coating to protect exposed area before cross-sectioning. The high resolution TEM micrographs were recorded at an accelerating voltage of 200 kV. Elemental mapping images were acquired by a dual energy dispersive X-ray spectroscopy (EDS) detectors system. The elements were first identified from the EDS spectrum and then selected for mapping.

**Device fabrication.** Contact electrodes were written on a cleaned Si/SiO$_2$ substrate using a standard electron beam lithography process, followed by the Cr/Au deposition using an e-beam evaporator. Typical thicknesses of Cr and Au are 2 and 25 nm, respectively. The pre-pattern electrode substrates were cleaned using Piranha method. BSTS were exfoliated into thin flakes on a PDMS and selected using an optical microscope. The BSTS flake was then transferred onto the pre-pattern electrodes using a home-made transfer stage. Similar steps were repeated for hBN and Gr flakes. Stamping of hBN flake onto the BSTS crystal was done at room temperature. The sample was then heated to 90ºC while transferring the top Gr layer. The heating step is critical to prevent the water vapor trapping between the hBN and Gr interface.

**Magneto-electrical transport measurements.** Room temperature electrical transport properties of the samples were characterized in a vacuum high precision probe station. Low temperature transport measurements were performed in a variable temperature insert (base temperature of 1.6 Kelvin) equipped with a superconducting magnetic up to 9 Tesla. Two synchronized Stanford Research SR830 lock-in amplifiers were used to measure the longitudinal and Hall resistances concurrently on BSTS device in a Hall bar configuration. The lock-in amplifiers were operated at a frequency of 17.777 Hz. The device is typically sourced with a constant a.c. excitation current of 10-100 nA. Keithley 2400 source meter was utilized to source a d.c. gate voltage to the Gr gate electrode.

**References**

1. Xia, Y. *et al.* Observation of a large-gap topological-insulator class with a single Dirac cone on the surface. *Nature Phys.*, **5**, 398-402 (2009).

2. Zhang, H. *et al.* Topological insulators in $Bi_2Se_3$, $Bi_2Te_3$ and $Sb_2Te_3$ with a single Dirac cone on the surface. *Nature Phys.*, **5**, 438-442 (2009).

3. Arakane, T. *et al*. Tunable Dirac cone in the topological insulator $Bi_{2-x}Sb_xTe_{3-y}Se_y$. *Nat. Commun.*, **3**, 636, (2012).

4. Analytis, J. G. *et al.* Two-dimensional surface state in the quantum limit of a topological insulator. *Nat. Phys.*, **6**, 960-964 (2010).

5. Xu, Y. *et al.* Observation of topological surface state quantum Hall effect in an intrinsic three-dimensional topological insulator. *Nat. Phys.*, **10**, 956-963 (2014).

6. Yoshimi, R. *et al.* Quantum Hall effect on top and bottom surface states of topological insulator $(Bi_{1-x}Sb_x)_2Te_3$ films. *Nat. Commun.*, **6**, 6627 (2015).

7. Xu, Y., Miotkowski, I. & Chen, Y. P. Quantum transport of two-dimensional species Dirac fermions in dual-gated three-dimensional topological insulators. *Nature Comm.*, **7**, 11434 (2015).

8. Zhang, S. *et al.* Anomalous quantization trajectory and parity anomaly in Co cluster decorated $BiSbTeSe_2$ nanodevices. *Nature Comm.*, **8**, 977 (2017).

9. Li, C. *et al.* Interaction between counter-propagating quantum Hall edge channels in the 3D topological insulator BiSbTeSe$_2$. *Phys. Rev. B*, **96**, 195427 (2017).

10. Hasan, M. Z. & Kane, C. L. Colloquium: Topological insulators. *Rev. Mod. Phys.*, **82**, 3045 (2010).

11. Fatemi, V. *et al.* Electrostatic coupling between two surfaces of a topological insulator nanodevice. *Phys. Rev. Lett.*, **113**, 206801 (2014).

12. Yang, F. *et al.* Dual-gated topological insulator thin-film device for efficient fermi-level tuning. *ACS Nano*, **9**, 4050-4055 (2015).

13. Geim, A. K. & Grigorieva, I. V. Van der Waals heterostructures. *Nature*, **499**, 419-425 (2013).

14. Chen, J.-H., Jang, C., Xiao, S., Ishigami, M. & Fuhrer, M. S. Intrinsic and extrinsic performance limits of graphene devices on SiO$_2$. *Nature Nanotechnol.*, **3**, 206-209 (2008).

15. Dean, C. R. *et al.* Boron nitride substrates for high-quality graphene electronics. *Nat. Nanotechol.* **5**, 722-726 (2010).

16. Xue, J. *et al.* Scanning tunnelling microscopy and spectroscopy of ultra-flat graphene on hexagonal boron nitride. *Nat. Mater.* **10**, 282-285 (2011).

17. Qi, X.L. & Zhang, S.C. Topological insulators and superconductors. *Rev. Mod. Phys.*, **83**, 1057 (2011).

18. Alegria, L. D. *et al.* large anomalous Hall effect in ferromagnetic insulator-topological insulator heterostructures. *Appl. Phys. Lett.*, **105**, 053512 (2014).

19. Zibrov, A. A. *et al.* Tunable interacting composite fermion phases in a half-filled bilayer-graphene Landau level. *Nature* **549**, 360-364 (2017).

20. Khmelnitskii, D. E. Quantization of Hall conductivity. *JETP Lett.*, **38**, 552 (1983).

21. Sessi, P. *et al.* Dual nature of magnetic dopants and competing trends in topological insulators. *Nat. Comm.*, **7**, 12027 (2016).

22. Jiang, Y. *et al.* Mass acquisition of Dirac fermions in magnetically doped topological insulator $Sb_2Te_3$ films. *Phys. Rev. B*, **92**, 195418 (2015).

23. Yoshimi, R. *et al.* Quantum Hall states stabilized in semi-magnetic bilayers of topological insulators. *Nat. Comm.*, **6**, 8530 (2015).

24. Zhang, Z. *et al.* Zeeman effect of the topological surface states revealed by quantum oscillations up to 91 Tesla. *Phys. Rev. B*, **92**, 235402 (2015).

25. Chang, C.-Z. *et al.* Zero-field dissipationless chiral edge transport and the nature of dissipation in the quantum anomalous Hall state. *Phys. Rev. Lett.* **115**, 057206 (2015).

26. Han, K. B. *et al.* Enhancement in surface mobility and quantum transport of $Bi_{2-x}Sb_xTe_{3-y}Se_y$ topological insulator by controlling the crystal growth conditions. arXiv:1805.09226 (2017).

27. Zhang, X. *et al.* Magnetic anisotropy of the single-crystalline ferromagnetic insulator $Cr_2Ge_2Te_6$. *Japan. J. Appl. Phys.*, **55**, 033001 (2016).

**Acknowledgements**

This work was supported by the NSF MRSEC program at the University of Utah under grant # DMR 1121252. Work done at SKKU was supported by an Institute for Information & Communications Technology Promotion (IITP) grant (B0117-16-1003). The authors acknowledge Ryan McLaughlin and Shuwan Liu for their assistance in Sagnac interferometer measurements.

**Author contributions**

**Additional information**

**Supplementary Information** accompanies this paper at http://www.nature.com/naturecommunications

**Competing financial interests:** The authors declare no competing financial interests.

**Figures**

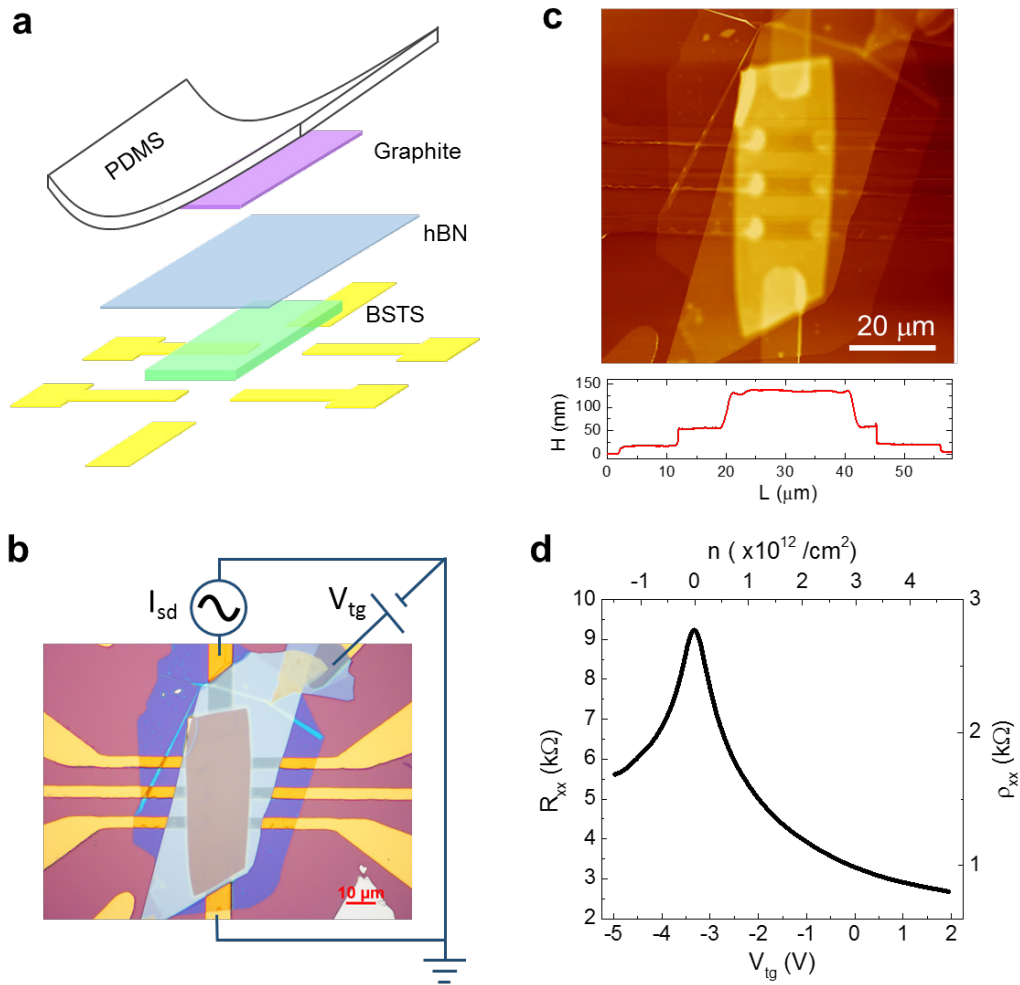

**Figure 1 Device fabrication.** (**a**) Schematic of the mechanical transfer of two-dimensional materials on pre-patterned electrodes. (**b**) Optical image of a transferred BSTS/hBN/Gr device with the topgate measurement scheme. (**c**) Atomic force microscopy image of the BSTS/hBN/Gr stack device and the height profile along the horizontal axis. (**d**) Four-probe resistance and resistivity as a function of topgate voltage applied between BSTS and graphite layers measured at cryogenic temperature of 1.6 K.

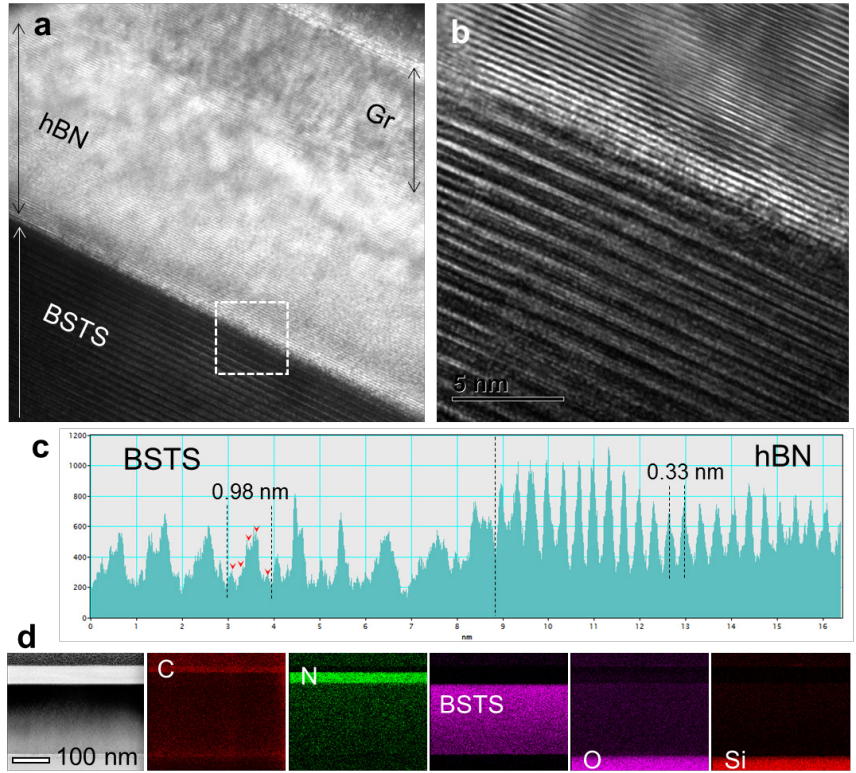

**Figure 2 Transmission electron microscopy (TEM) analysis.** (**a**) Cross-section high resolution TEM micrograph of a mechanically transferred BSTS/hBN/Gr stack. (**b**) Magnified TEM micrograph at the BSTS and hBN interface (marked as white frame). (**c**) Contrast line profile along the perpendicular plane across the BSTS and hBN interface. (**d**) Energy dispersive X-ray (EDX) mapping images of the BSTS/hBN/Gr stack.

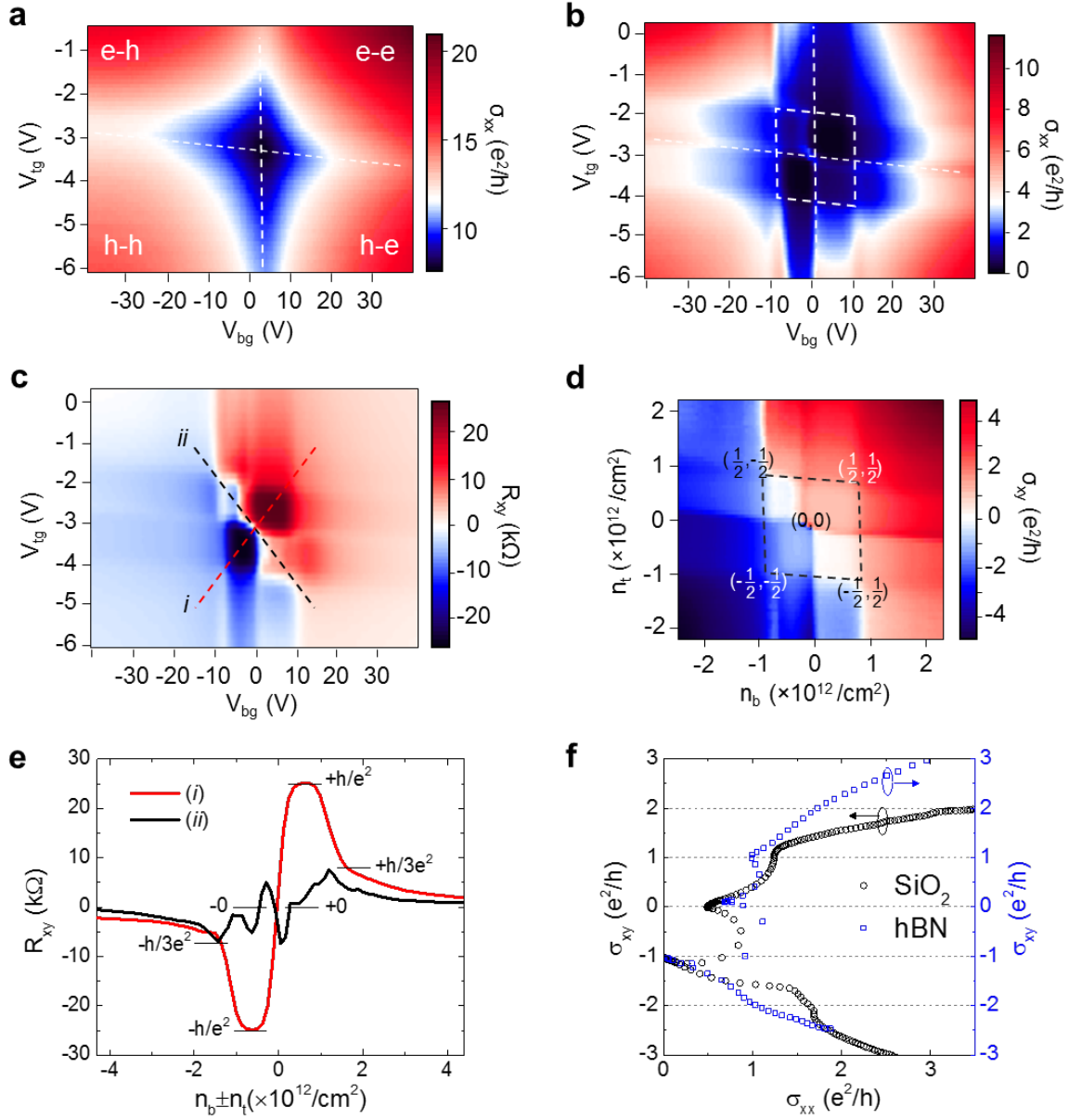

**Figure 3 Quantum transport of the BSTS/hBN/Gr dual-gated device.** Two-dimensional map of the longitudinal conductivity ($\sigma_{xx}$) of the BSTS as a function of topgate and backgate voltages at magnetic field of (**a**) 0 T, and (**b**) 9 T at cryogenic temperature of 1.6 K. (**c**) Hall resistance ($R_{xy}$) of the BSTS as a function of topgate and backgate voltages at magnetic field of 9 T at temperature of 1.6 K. (**d**) Hall conductivity ($\sigma_{xy}$) map of the BSTS as a function of charge densities induced at the top ($n_t$) and bottom ($n_b$) surfaces. (**e**) Line profiles of the Hall resistances extracted at (*i*) $n_b-n_t=0$ and (*ii*) $n_b+n_t=0$. (**f**) Renormalization group flow diagram ($\sigma_{xy}$ versus $\sigma_{xx}$) plots of the BSTS gated by hBN (top gate) and SiO$_2$ (bottom gate) dielectric layers.

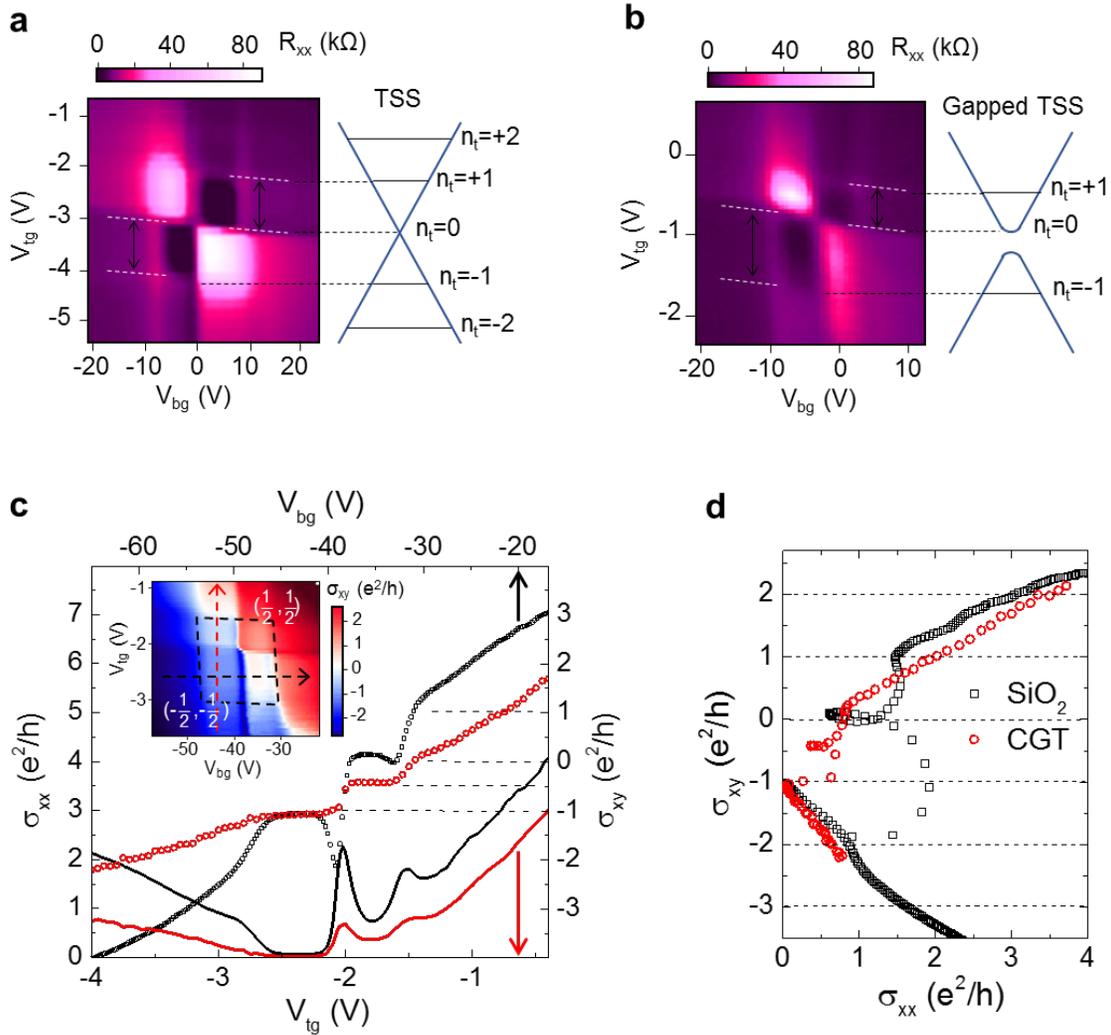

**Figure 4 Gating effect and magnetoelectric transport of the BSTS/CGT/Gr device.** 2D maps of the longitudinal resistance as a function of the $V_{tg}$ and $V_{bg}$ at magnetic field of 9 T for (**a**) BSTS/hBN/Gr and (**b**) BSTS/CGT/Gr devices. Schematic of the LL band diagrams of the top gapless topological surface state (TSS) and gapped TSS corresponding to the longitudinal resistance maps in (a) and (b), respectively, are shown on the right side. (**c**) Longitudinal and Hall conductivities of the BSTS as a function of topgate (backgate) voltages at $V_{bg}$= -43V ($V_{tg}$= -2.7V) and magnetic field of 9 T. Inset in (c) is the map of the Hall conductivity of the BSTS as the function of dual-gate voltages. (**d**) Renormalization group flow plots of the BSTS gated by CGT and $SiO_2$ dielectric layers.